\documentclass[amsfonts,amsmath,prd,preprint,nofootinbib]{revtex4}
\newcommand{\beq}{\begin{equation}}
\newcommand{\eeq}{\end{equation}}

\newcommand{\bsp}{\begin{split}}

\usepackage{epsfig,bbm,cancel,ulem}
\usepackage[breaklinks=true]{hyperref}
\graphicspath{{./figures/}}

\newcommand{\mc}[1]{\mathcal{#1}}

\begin{document}

\title{Gravitational field of global monopole within the Eddington-inspired Born-Infeld theory of gravity}

\author{Reyhan D. Lambaga}
\email{reyhan.dani@sci.ui.ac.id}
\author{Handhika S. Ramadhan}
\email{hramad@ui.ac.id}
\affiliation{Departemen Fisika, FMIPA, Universitas Indonesia, Depok, 16424, Indonesia. }
\def\changenote#1{\footnote{\bf #1}}

\begin{abstract}
Within the framework of the recent Eddington-inspired Born-Infeld (EiBI) theory we study gravitational field around an $SO(3)$ global monopole. The solution also suffers from the deficit solid angle as in the Barriola-Vilenkin metric but shows a distinct feature that cannot be transformed away unless in the vanishing EiBI coupling constant, $\kappa$. When seen as a black hole eating up a global monopole, the corresponding Schwarzschild horizon is shrunk by $\kappa$. The deficit solid angle makes the space is globally not Euclidean, and to first order in $\kappa$ (weak-field limit)  the deflection angle of light is smaller than its Barriola-Vilenkin counterpart.
 
\end{abstract}

\maketitle
\thispagestyle{empty}
\setcounter{page}{1}

\section{Introduction} \label{sec:introduction}

Global monopole is an exotic object that may have been formed during the phase transition in the very early universe when the corresponding vacuum is non-contractible, ${\cal M}\cong S^2$~\cite{vilenkinshellard}. Such object can exist, for example, due to the spontaneous breaking of global $SO(3)$ symmetry. Its gravitational field has been studied by Barriola and Vilenkin (BV), who found that the metric is Minkowski-like, though not flat, and suffers from deficit solid angle~\cite{Barriola:1989hx}. When viewed as a metric around a massive monopole, within a reasonable astrophysical scale the mass $M$ is tiny. Physically, this means that the monopole effectively exerts no gravitational force on its surrounding.  Further investigation by Harari and Lousto shows that this tiny mass is in fact negative~\cite{Harari:1990cz}. It implies that if we insist on calculating the gravitational force, it is repulsive. On the other hand, when the monopole core is much smaller than its corresponding Schwarzschild radius, $\delta\ll GM$, the solution describes the collapse of a star near a massive global monopole~\cite{Carames:2017ngt}; that is, a (non-asymptotically flat) black hole with scalar hair~\cite{Dadhich:1997mh}.

Inspired by the BV solution, in recent years global monopole in ``modified environment" has extensively been studied. By ``modified environment" we mean that either: (i) the energy-momentum tensor is non-canonical, or (ii) the gravity part is non-GR (General Relativity). The former deals with modification of the Higgs lagrangian, mostly with non-canonical kinetic term~\cite{Jin:2007fz, Liu:2009eh, Prasetyo:2017rij}. The latter looks for spacetime geometry around global monopole, mostly in the framework of $f(R)$ theories of gravity~\cite{Carames:2017ngt, Carames:2011xi, Carames:2011uu}. In this $f(R)$ gravity framework, thermodynamical properties of static black hole with global monopole are investigated in~\cite{Man:2013sf, Lustosa:2015hwa}, solutions for the corresponding rotating black hole is studied in~\cite{MoraisGraca:2012sq}, while astrophysical signature (for example, strong gravitational lensing) of such massive monopole is discussed in~\cite{Man:2012fa}. 

Recently an old proposal by Eddington for gravitational theory has been resurrected by incorporating the Born-Infeld structure into the action~\cite{Banados:2010ix}. The so-called Eddington-inspired Born-Infeld (EiBI) theory of gravity is shown to be equivalent to the ordinary Einstein's General Relativity in the vacuum, while gives distinct features when matters are included. From theoretical point of view, EiBI theory enjoys internal consistency since it is free of instabilities and ghosts~\cite{Delsate:2012ky}. As a scientific proposal, perhaps one of its most interesting predictive power is that there is cosmologically no singularity; that the universe is born out of a regular bounce, and that the gravitational collapse leads to entirely singularity-free state~\cite{Banados:2010ix} (however, see~\cite{Pani:2012qd}). In the context of nuclear astrophysics, EiBI has been a new favorite model of modified gravity to study neutron stars (NS), or the other way around. In order not to obstruct the success of GR's prediction for astrophysical phenomena in lower energy scale, the value of $\kappa$ can be constrained by observations of the radii of low mass NS~\cite{Sotani:2014goa}. In particular, it was shown in~\cite{Avelino:2012ge} that a stable NS can be held together by its own gravity should $\kappa<10^{-2}\ \rm{m^5kg^{-1}s^{-2}}$. An even stronger constraint comes from the compatibility of maximum observed mass of NS and the absence of ``hyperon puzzle", given by\footnote{See also~\cite{Prasetyo:2017hrb} for the compatibility of NS within a braneworld scenario in the framework of EiBI theory.} $2.6\times 10^{-5}~\rm{m^5 kg^{-1} s^{-2}}~\lesssim~\kappa \lesssim~4\times 10^{-5}~\rm{m^5 kg^{-1} s^{-2}}$~\cite{Qauli:2016vza, Qauli:2017ntr}. However, this stringent constraint works only for NS. For the black hole the value of $\kappa$ can be larger. For a comprehensive review on EiBI gravity, see~\cite{BeltranJimenez:2017doy} and the references therein.

While the simplest case of black hole in EiBI theory is the electrically-charged one studied in~\cite{Sotani:2014lua, Wei:2014dka}, to the best of our knowledge there is yet any study on the EiBi black hole with global monopole. It is therefore the aim of this work. In this paper we systematically investigate the properties of gravitational field of global monopole in EiBI theory along with its corresponding black hole. This paper is therefore organized as follows. In Section \ref{sec:EiBIGoldstone} we present the solutions of gravitating Goldstone field in EiBI gravity, which we interpret in Section \ref{sec:Gravfield} as a metric outside a global monopole. Section \ref{sec:geodesic} is devoted to the analysis of geodesic of test (massive or massless) particles around it, and in addition, the deflection of light due to this monopole black hole. Finally we summarize our analysis in Section \ref{sec:conclusions}.

\section{The EiBI-Goldstone Model} \label{sec:EiBIGoldstone}

The action is~\cite{Banados:2010ix, Sotani:2014lua}:
\beq
S = \frac{1}{8\pi \kappa}\int d^4x \left(\sqrt{\lvert g_{\mu \nu} + \kappa R_{\mu \nu}\rvert} - \lambda \sqrt{-g} \right) + S_{M} \left[g, \Phi_{M}\right],\label{eq:action}
\eeq
where $\lambda\equiv1+\kappa\Lambda$, and $\kappa$ is the so-called Eddington parameter.  In the vacuum limit, it can be shown that the EiBI theory is identical to the Einstein's GR. The matter lagrangian is given by the global $SO(3)$ Lagrangian:
\beq
{\mathcal L}={1\over2}\partial_{\mu}\Phi^a\partial^{\mu}\Phi^a-{\sigma\over4}\left(\Phi^a\Phi^a-\eta^2\right)^2,
\eeq
with $a=1, 2, 3$. This theory is known to possess global monopole solution.

As shown in~\cite{Deser:1998rj}, varying the action~\eqref{eq:action} with respect to the metric field will lead to fourth-order field equations with ghosts. In order to avoid it, Vollick invokes the Palatini formalism~\cite{Vollick:2003qp}; {\it i.e.}, we treat $\Gamma^{\mu}_{\alpha\beta}$ and $g_{\mu\nu}$ as two independent fields, and the Ricci tensor $R_{\mu\nu}$ depends on $\Gamma^{\mu}_{\alpha\beta}$, not on $g_{\mu\nu}$. We assume the spacetime manifold to be torsion-less; that is, $\Gamma^{\mu}_{\alpha\beta}=\Gamma^{\mu}_{\beta\alpha}$. Thus, only the symmetric part of the Ricci tensor is considered. This enforces the projective invariance to become the symmetry of full the theory~\cite{BeltranJimenez:2017doy}. In this formalism the field equations are:
\beq
\label{eq:field1}
q_{\mu \nu} = g_{\mu\nu} + \kappa R_{\mu \nu},
\eeq
\beq
\label{eq:field2}
\sqrt{-q}q^{\mu\nu} = \lambda\sqrt{-g}g^{\mu\nu} - 8\pi\kappa\sqrt{-g}T^{\mu\nu},
\eeq
with
\beq
\label{eq:constraintgamma}
\Gamma^{\mu}_{\alpha \beta} = \frac{1}{2} q^{\mu \nu} \left(q_{\nu\alpha , \beta}+q_{\nu\beta , \alpha}- q_{\alpha \beta ,\nu}\right),
\eeq
where $q_{\mu\nu}$ is the {\it auxiliary} metric. The energy-momentum tensor $T_{\mu\nu}$ depends on $g_{\mu\nu}$.

Taking the spherically-symmetric ansatz for the metric:
\beq
g_{\mu\nu}\mathrm{d}x^{\mu}\mathrm{d}x^{\nu} = -A^2 f dt^2 + f^{-1} dr^2 + r^2 d\Omega^2,
\eeq
\beq
q_{\mu\nu}dx^{\mu}dx^{\nu} = -G^2 F dt^2 + F^{-1} dr^2 + H^2 d\Omega^2,
\eeq
and hedgehog configuration for the scalar fields:
\beq
\label{eq:hedgehog}
\Phi^a=\eta \phi(r){x^a\over r},
\eeq
Eq.~\eqref{eq:field1} reduces to
\begin{eqnarray}
{2\over\kappa F}\left({A^2 f\over G^2F} - 1\right)&=&\frac{F''}{F} + \frac{2G''}{G} + \frac{3G'F}{GF}'+\frac{2F'H'}{FH}+\frac{4G'H'}{GH}, \label{dua0}\\
{2\over\kappa F}\left({F\over f} - 1\right)&=&\frac{F''}{F}+\frac{2G''}{G}+\frac{4H''}{H}+\frac{2F'H'}{FH}+\frac{3F'G'}{FG}, \label{dua1}\\
\frac{1}{\kappa F}\left(\frac{r^2}{H^2} - 1\right)&=&-\frac{1}{H^2F} +\frac{F'H'}{FH}+\frac{H'^2}{H^2} + \frac{H''}{H} + \frac{H'G'}{HG} \label{dua2}.
\end{eqnarray}

The hedgehog ansatz~\eqref{eq:hedgehog} yields the following exterior energy-momentum tensors are (taking\footnote{This approximation can be made ``exact" by looking from the point of view of non-linear $\sigma$ model coupled to EiBI. The hedgehog ansatz for the scalar field is an exact solution of the corresponding $\sigma$ model equation. See Refs.~\cite{Prasetyo:2017rij, Tan:2017egu}.} $\phi\approx1$):
\beq
T^r_r = T^t_t = -\frac{\eta^2}{r^2}
\eeq
\beq
T^\theta_\theta = T^\psi_\psi = 0. 
\eeq
When substituted to Eq.~\eqref{eq:field2} they give
\begin{eqnarray}
\frac{A H^2 f}{G r^2 F}&=&\lambda + 8\pi \kappa   \frac{\eta^2}{r^2},\label{tiga1}\\
\frac{G H^2 F}{A r^2 f}&=&\lambda + 8\pi \kappa   \frac{\eta^2}{r^2},\label{tiga2}\\
G&=&\lambda A.\label{tiga3}
\end{eqnarray}

It is elementary to solve \eqref{tiga1}-\eqref{tiga3} to obtain:
\begin{eqnarray}
H^2&=&\lambda r^2 + 8 \pi \kappa \eta^2,\label{hasiltiga1}\\
G&=&\lambda A,\label{hasiltiga2}\\
F&=&\frac{f}{\lambda}.\label{hasiltiga3}
\end{eqnarray}

Eqs.~\eqref{hasiltiga2} and \eqref{hasiltiga3} can be combined to yield $GF=Af$. This, when inserted into~\eqref{dua0}-\eqref{dua1} along with \eqref{hasiltiga1}, yields
\begin{equation}
\frac{H''}{G} - \frac{G'H'}{G^2}=0\ \ \ \rightarrow \ \left(\frac{H'}{G}\right)'=0.
\end{equation}
The solution is simple,
\begin{equation}
\label{nilaiG}
G=c_1 H', 
\end{equation}
with $c_1$ an integration constant. We can thus determine $A$:
\begin{equation}
A=\frac{c_1 r}{\sqrt{\lambda r^2 + 8 \pi \kappa \eta^2}}.
\end{equation}
As $\eta \rightarrow 0$, the solution should return to Schwarzschild's. This condition can only be achieved with $A = 1$, thus $c_1 = \sqrt{\lambda}$.

Inserting~\eqref{nilaiG} into \eqref{dua2}, we obtain
\begin{equation}
\frac{\lambda r}{\kappa \sqrt{\lambda r^2 + 8\pi \kappa \eta^2}}\left(\kappa + r^2 (1-\lambda) - 8\pi \kappa \eta^2 \right)=\left(\frac{f\lambda r^2}{ \sqrt{\lambda r^2 + 8\pi \kappa \eta^2}}\right)'.
\end{equation}
The metric $f$ can therefore be obtained as
\beq
f=\frac{\sqrt{\lambda r^2 + 8\pi \kappa \eta^2}}{\lambda r^2 }\int \frac{\lambda r}{\sqrt{\lambda r^2 + 8\pi \kappa \eta^2}}\left(\frac{r^2}{\kappa} (1-\lambda) + 1 - 8\pi \eta^2 \right) \mathrm{d}r.
\eeq
This integral is elementary. Defining 
\begin{equation}
v\equiv\lambda r^2 + 8\pi \kappa \eta^2,
\end{equation}
we obtain
\beq
f=\frac{v^2}{3\lambda^2 \kappa r^2}(1-\lambda) - \frac{v \Delta}{\lambda^2 r^2} + \frac{v}{\lambda r^2} + \frac{\sqrt{v}}{\lambda r^2 }C,
\eeq
with $\Delta\equiv8\pi\eta^2$. In the limit $\eta\rightarrow0$ the solution should reduce to Schwarzschild-de Sitter's,
\beq
f=1-\frac{r^2 \Lambda}{3} + \frac{C}{\sqrt{\lambda}r}.  
\eeq
Therefore $C = -2M\sqrt{\lambda}$, where $M$ can be identified as the mass of the monopole.

Thus, the metric solutions are given by,
\begin{eqnarray}
g_{\mu\nu}\mathrm{d}x^{\mu}\mathrm{d}x^{\nu}&=&-\left(1 - \frac{\Delta}{\lambda} -\frac{v \Lambda}{3\lambda}- \frac{2M\sqrt{\lambda}}{\sqrt{v}}\right)\mathrm{d}t^2 +\frac{\lambda r^2}{v}\left(1 - \frac{\Delta}{\lambda} -\frac{v \Lambda}{3\lambda}- \frac{2M\sqrt{\lambda}}{\sqrt{v}}\right)^{-1}\mathrm{d}r^2\nonumber\\ 
&&+ r^2\mathrm{d}\Omega^2,\label{eq:gsol}
\end{eqnarray}
and
\begin{eqnarray}
q_{\mu\nu}\mathrm{d}x^{\mu}\mathrm{d}x^{\nu}&=&-\lambda\left(1 - \frac{\Delta}{\lambda} -\frac{v \Lambda}{3\lambda}- \frac{2M\sqrt{\lambda}}{\sqrt{v}}\right)\mathrm{d}t^2 +\frac{\lambda^2 r^2}{v}\left(1 - \frac{\Delta}{\lambda} -\frac{v \Lambda}{3\lambda}- \frac{2M\sqrt{\lambda}}{\sqrt{v}}\right)^{-1}\mathrm{d}r^2\nonumber\\ 
&&+ v\mathrm{d}\Omega^2.\label{eq:qsol}
\end{eqnarray}
These are our main results. Note that what matters physically is the physical metric $g_{\mu\nu}$. In the following sections, we shall examine the significance of this genuine solution with $\Lambda = 0$.

\section{Gravitational Field of EiBI Global Monopole} \label{sec:Gravfield}

\begin{figure}[h]
	\includegraphics[scale=0.5]{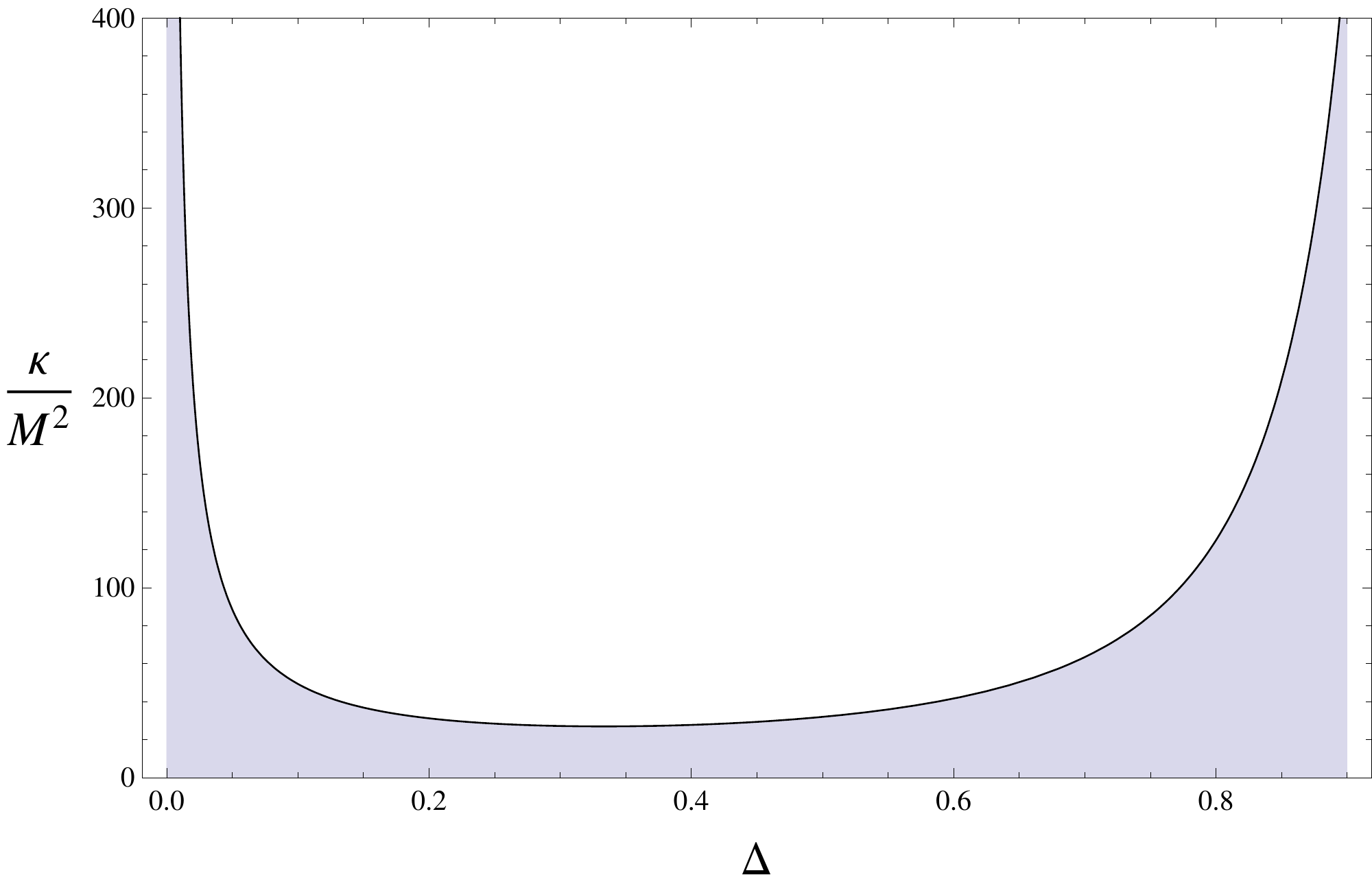}
	\caption{The allowed value of $\kappa/M^2$, indicated by gray area below the curve.}
	\label{fig:param}
\end{figure}

For simplicity, through the rest of this work we shall work in $\lambda=1$ (or equivalent to $\Lambda=0$) condition,
\begin{eqnarray}
\label{metricsolution}
ds^2= -\left(1 -\Delta - \frac{2M}{\sqrt{r^2 + \Delta \kappa}}\right)\mathrm{d}t^2 +\frac{r^2}{r^2 + \Delta \kappa}\left(1 - \Delta  - \frac{2M}{\sqrt{r^2 + \Delta \kappa}}\right)^{-1}\mathrm{d}r^2+ r^2\mathrm{d}\Omega^2.
\end{eqnarray}
 Rescaling the metric $t \rightarrow t\sqrt{1-\Delta}$ and $r \rightarrow r\left(\sqrt{1-\Delta}\right)^{-1}$, simultaneously also $M\rightarrow M \left(1-\Delta\right)^{-3/2}$ and $\kappa\rightarrow\Delta\kappa\left(1-\Delta\right)^{-1}$, we obtain
\begin{eqnarray}
\label{metricsolutiondeficitangle}
ds^2=-\left(1-\frac{2M}{\sqrt{r^2 + \kappa}}\right)dt^2 +\frac{r^2}{r^2 +\kappa}\left(1-\frac{2M}{\sqrt{r^2 + \kappa}}\right)^{-1}dr^2+ r^2\left(1-\Delta\right)d\Omega^2. 
\end{eqnarray}
We can see that the solution is distinct from BV's\footnote{Obviously we can further rescale $r \rightarrow \sqrt{r^2 +\kappa}$ such that the metric now takes the form 
\beq
ds^2=-\left(1-\frac{2M}{r}\right)dt^2 +\left(1-\frac{2M}{r}\right)^{-1}dr^2+ r^2\left(1-\Delta\left(1+{\kappa\over r^2}\right)\right)d\Omega^2,\nonumber
\eeq
where it differs from the ordinary BV in the existence of singularity at $r=\sqrt{\Delta\kappa\over1-\Delta}$, apart from the one at the origin. This singularity in the solid angle resembles the charged black hole solutions in low-dimensional string theory~\cite{Garfinkle:1990qj}. However, there is no new information here, since the transformation only shifts the positions of singularity. It is also worth pointing out that the auxiliary metric \eqref{eq:qsol} becomes, under the same rescaling,
\begin{equation}
ds^2 =-\left(1 - \frac{2M}{r}\right)dt^2 +\left(1 - \frac{2M}{r}\right)^{-1}dr^2 + r^2(1-\Delta)d\Omega^2,\nonumber
\end{equation}
which exactly resembles BV's.}, and can only reduce to them in the limit $\kappa\rightarrow0$. Eq.~\eqref{metricsolution} (and \eqref{metricsolutiondeficitangle}) describe a black hole with global monopole. The shared feature with BV solution is that they are both not asymptotically-flat. There is a deficit solid angle $\Delta$. As $\eta$ exceeds some critical value, $\eta_{crit}\equiv1/\sqrt{8\pi}$, the solid angle disappears and the spacetime develops conical singularity. There is also an apparent singularity at
\beq
r=\sqrt{{4M^2\over\left(1-\Delta\right)^2}-\Delta\kappa}.
\eeq
This is a horizon whose value differs from the Schwarzschild by the appearance of $\kappa$. It appears that the modification of gravity does not modify the number of horizons. In some sense, the {\it black-hole-has-no-scalar-hair} theorem still works in this theory. In order to avoid naked singularity, the value of $\kappa$ is constrained to 
\begin{equation}
\kappa/M^2\leq{4\over\Delta\left(1-\Delta\right)^2}.
\end{equation}
This gives us the range of possible value of $\kappa$, which we plot in Fig.~\ref{fig:param}.

The upper bound for the dimensionless quantity $\kappa/M^2$ is given by~\cite{Sotani:2014lua},
\begin{equation}
\kappa/M^2\lesssim6.87\times10^3\times\left({M_\odot\over M}\right)^2,
\end{equation}
where $M_\odot$ is the sun mass and $M$ is the mass of the corresponding black hole. 

The black hole interpretation above works whenever the corresponding Schwarzschild radius is much greater than the size of monopole's core, $M\gg\delta$. In asymptotically-flat space, the typical core is of the order $\delta\sim\eta^{-1}$~\cite{Barriola:1989hx}. When the mass is negligible, $\delta\gg M$ the metric~\eqref{eq:gsol} describes gravitational field around a massive monopole. In this case, the spacetime yields
\beq
\label{metricm0}
ds^2=-dt^2 +\frac{r^2}{r^2 +\kappa}dr^2+ r^2\left(1-\Delta\right)d\Omega^2. 
\eeq 
As in BV solution, this metric is locally non-flat. This can be seen from the corresponding Kretschmann scalar
\begin{eqnarray}
K&=& \frac{12 \kappa ^2}{r^8}-\frac{8 \Delta  \kappa }{(1-\Delta ) r^6}+\frac{4 \Delta ^2}{(1-\Delta )^2 r^4}.
\end{eqnarray}
Apart from $\Delta$, it is also $\kappa$ that gives additional contribution to the singularity at the origin. Note that this scalar is built from the physical metric $g$, not the auxiliary $q$. Although the Ricci tensor depends not on $g$, we nevertheless can construct ``Riemann curvature" of the physical $g$ metric, as was done in~\cite{Pani:2012qd} for the ``Ricci curvature".

Since solution~\eqref{metricm0} cannot be further transformed into the BV metric, they both describe two different inequivalent solutions. The modification of gravity gives correction to the BV solution, and these signatures may be observed cosmologically. In the following, we shall discuss the geodesic of massive particles and deflection of light around a (very massive) monopole.

\section{Geodesic equation and test particles around the monopole black hole} \label{sec:geodesic}

As is well-known, one peculiar feature of global monopole is that the gravitational field it exerts is zero (neglecting the tiny negative mass at the origin). This is also the case for our solution, as can be seen from~\eqref{metricm0}. In the weak-field regime (small $\kappa$) the Newtonian potential gravity is zero. Nevertheless, the deficit solid angle makes the global geometry not asymptotically-flat. 

On the other hand, a very massive monopole (or a black hole with global monopole) may affect the surrounding particles through its gravitational field. Here we study the classical motion of massive and massless test particle around such a black hole. We follow~\cite{Sotani:2014lua, Carames:2011xi} for the general formalism. The Langrangian for a test particle around this black hole is
\begin{eqnarray}
\mc{L} &=& \frac{1}{2} g_{\mu\nu}\frac{\mathrm{d}x^\mu}{\mathrm{d}\tau}\frac{\mathrm{d}x^\nu}{\mathrm{d}\tau}\\
&=& \frac{1}{2}\left(-B\dot{t}^2 +\frac{C}{B}\dot{r}^2 + r^2\dot{\theta}^2 + r^2 \sin{\theta}^2\dot{\psi}^2 \right) ,
\end{eqnarray}
where we define \(B \equiv \left(1 - \Delta - 2M/\sqrt{v}\right) \) and \(C \equiv r^2/v \). The dot here denotes the derivative with respect to \(\tau\).

The canonical momenta, \(p_\alpha = \frac{\partial \mathcal{L}}{\partial \dot{x}^\alpha}\), for this langrangian are,
\begin{eqnarray}
&p_t &= -B \dot{t} = -E, \\ 
&p_\phi &= r^2 \dot{\phi} = l, \label{eq:pl}
\end{eqnarray}
where $E$ and $l$ are constants that represent energy and angular momentum per unit rest mass, respectively.
 
Without loss of generality we may assume that the motion lies on a plane, thus we set $\theta = \pi /2$. Since the lagrangian is conserved along the geodesic, the following relation holds
\begin{equation}\label{eq:geoconv}
g_{\mu\nu} \frac{\mathrm{d}x^\mu}{\mathrm{d}\tau} \frac{\mathrm{d}x^\nu}{\mathrm{d}\tau} = - \epsilon ,
\end{equation}
where $\epsilon = 1$ and $0$ for massive and massless particles, respectively. The equation thus reduces to the ``Newtonian" motion with an effective potential~\cite{proceding},
\begin{equation} \label{eq:potprob}
C \dot{r}^2 + V_{eff}(r) = E^2,
\end{equation}
with 
\begin{equation}
V_{eff}(r) = \left(\epsilon + \frac{l^2}{r^2}\right)\left(1 - \Delta - \frac{2M}{\sqrt{v}}\right).
\end{equation}
The circular orbit of a test particle can then be determined by setting $\dot{r}=0$. This amounts to having $V_{eff}(r)=E^2$.

\subsection{Massive Particle Motion}

For massive particle, the effective potential is given by
\begin{equation}
V_{eff}(r) = \left(1 + \frac{l^2}{r^2}\right)\left(1 - \Delta - \frac{2M}{\sqrt{v}}\right)
\end{equation}
As is known from analytical mechanics, the radius of the minimum represents the radius of the stable circular orbit (SCO) of the test particle, while the radius at the peak gives the unstable one (UCO, {\it unstable circular orbit}). They are determined by the conditions ${dV_{eff}\over dr}\bigg|_{r=R} = 0$ and ${d^2V_{eff}\over dr^2} > 0$ (for SCO) or ${d^2V_{eff}\over dr^2} < 0$ (for UCO).

\begin{figure}[h]
\includegraphics[scale=0.7]{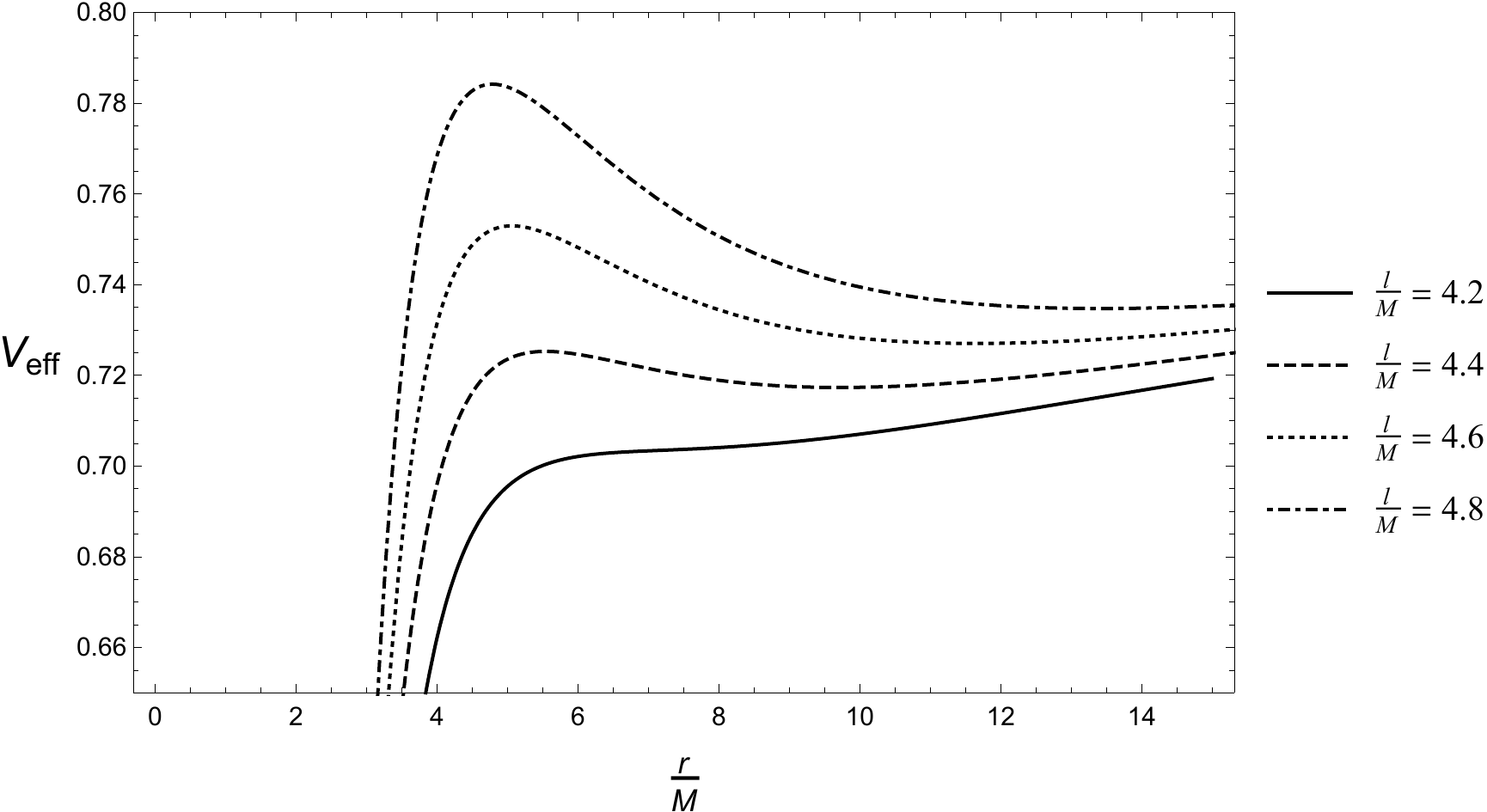}
\caption{$V_{eff}$ with various value of $\ell/M$ ($\kappa /M^2 = 5 $ and $ \Delta = 0.2$).}
\label{fig:VL}
\end{figure}

We plot the potential function for various conditions of the parameters. In Fig.~\ref{fig:VL}, we plot $V_{eff}$ with different value of $l/M$. As $l/M$ decreases, the maximum of effective potential decreases and the minimum point moves to the left.  As we decrease $l/M$ further, $R_{UCO}$ and $R_{SCO}$ merge. This overlap radius represents the {\it Innermost Stable Circular Orbit} (ISCO), $R_{ISCO}$~\cite{Sotani:2014lua}. Below this value the stable orbit disappears, and any particle that possesses these values shall collapse into the black hole.

\begin{figure}[h]
\includegraphics[scale=0.7]{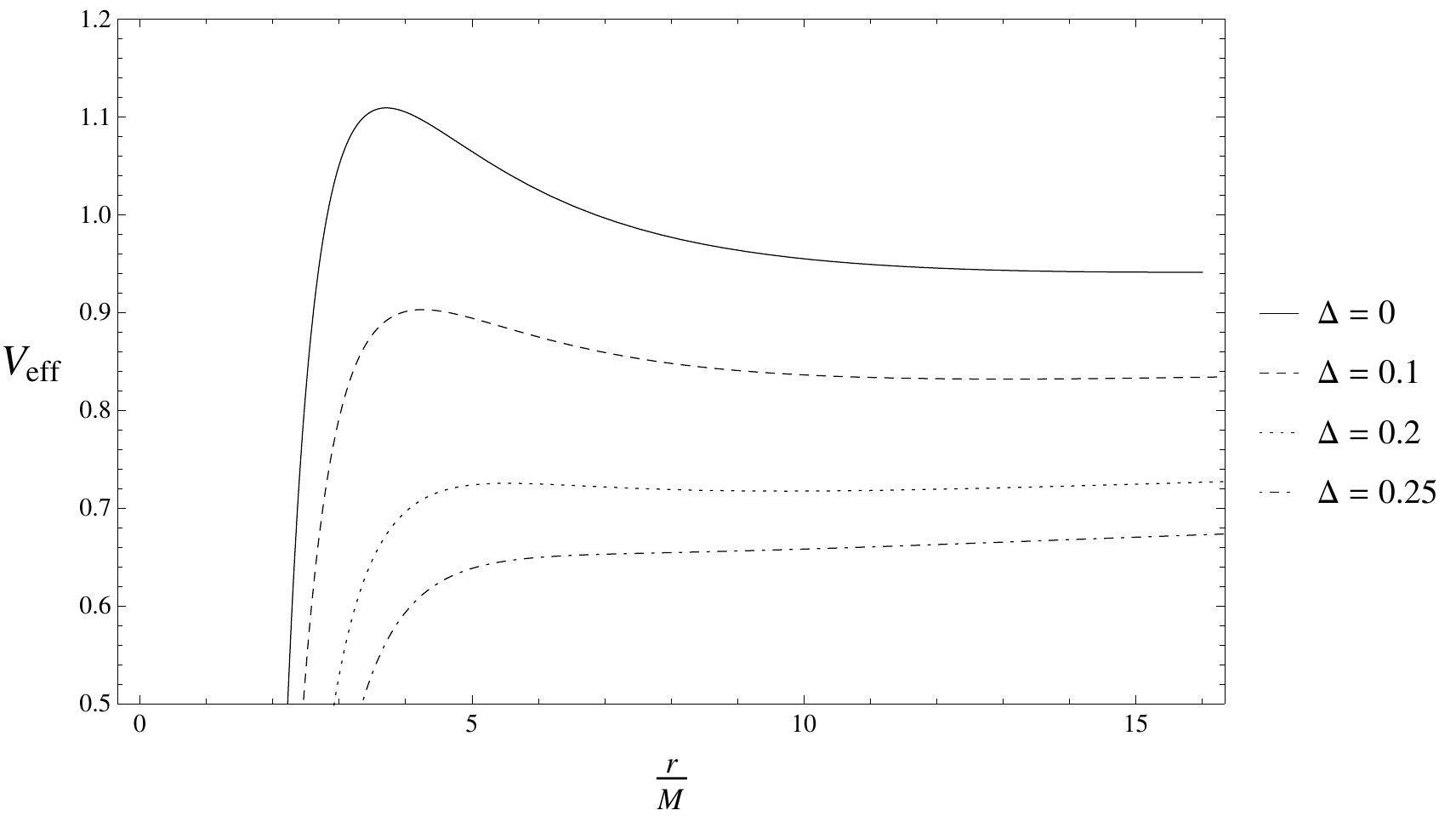}
\caption{$V_{eff}$ with various value of $\Delta$ ($\kappa /M^2 = 5 $ and $\ell/M = 4.4$).}
\label{fig:VD}
\end{figure}

A similar behavior is observed for the potential for variation of the monopole charge $\Delta$, as shown in Fig.~\ref{fig:VD}. The maximum of the potential decreases and the minimum point moves to the left as $\Delta$ increases. There also exists a maximum value of $\Delta$ which gives radius of ISCO. As we add more monopole charge, $R_{ISCO}$ disappears. 

\begin{figure}[h]
\includegraphics[scale=0.7]{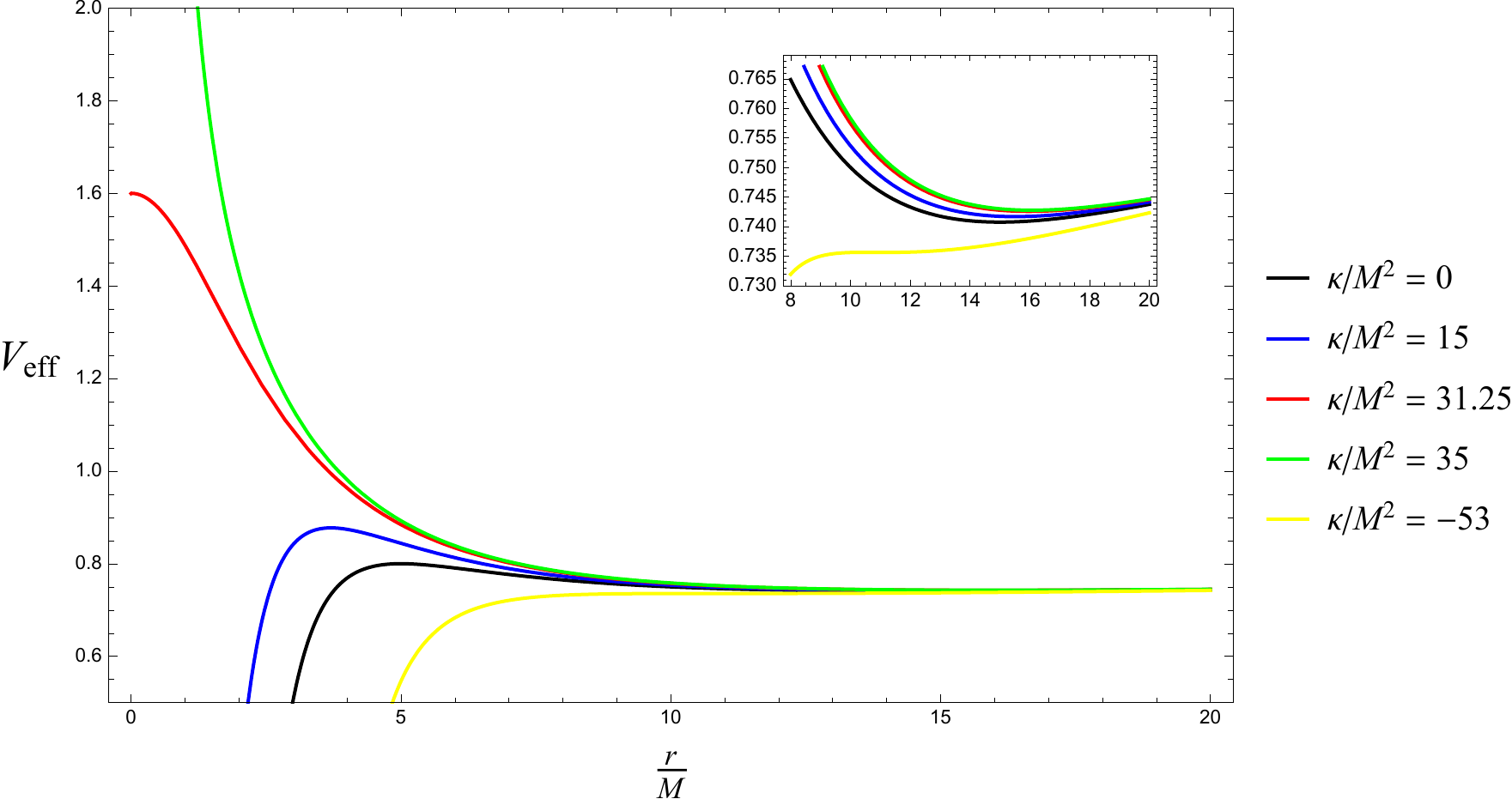}
\caption{$V_{eff}$ with various value of $\kappa /M^2$ ($\Delta = 0.2 $ and $\ell/M = 5$).}
\label{fig:kV}
\end{figure}

\begin{figure}[h]
	\includegraphics[scale=0.5]{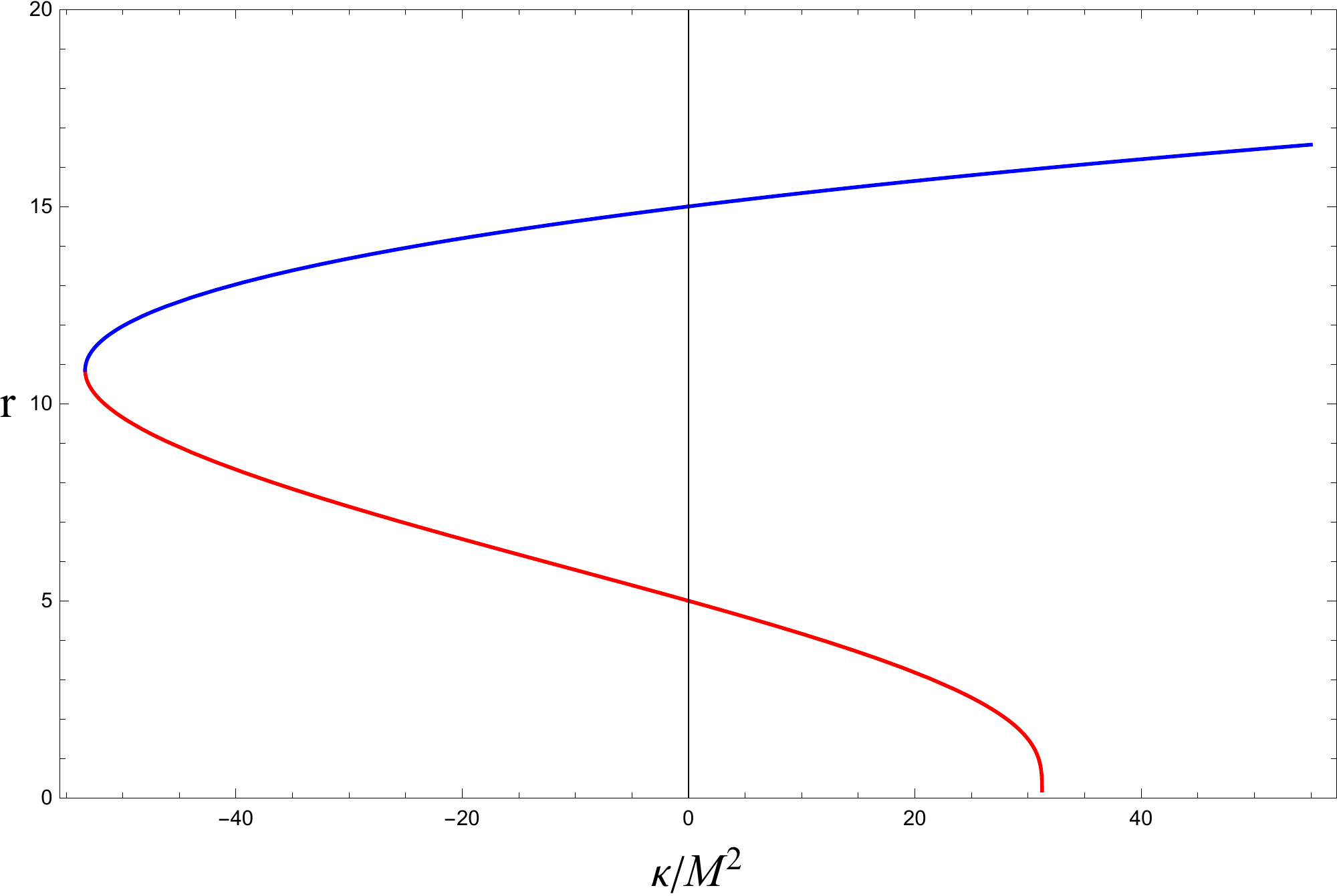}
	\caption{extrema of $V_{eff}$ as a function of $\kappa/M^2$ with $\Delta = 0.2 $ and $\ell/M = 5$.}
	\label{fig:Drk}
\end{figure}

A different behavior occurs when we vary $\kappa /M^2$, as we can see in Fig.~\ref{fig:kV}. As $\kappa /M^2$ decreases the peak of the potential decreases, while the $R_{SCO}$ moves to the left. This behavior continues even for negative $\kappa/M^2$. As we continue to decreases $\kappa/M^2$, there will be lower bound value for $\kappa /M^2$ which gives $R_{ISCO}$. Otherwise should we increase $\kappa /M^2$, the $R_{UCO}$ will move to the left. At some value of $\kappa /M^2$, this maximum point will reach $r = 0$. When this happen, $V_{eff}$ will have some finite value at the origin. Increasing $\kappa /M^2$ further, $V_{eff} \rightarrow + \infty$ as $r \rightarrow 0$. This behavior can be explained by plotting $R$ as the solutions of $\partial_r V_{eff} = 0$, shown in Fig. \ref{fig:Drk}. The red line indicates the unstable solution $R_{UCO}$ while the blue line is $R_{SCO}$. The ISCO radius can be seen in the left part of the graph, where the blue and red lines join together. Finite $V_{eff}$ at the origin occurs when the red line reaches $r = 0$. On the other hand, $V_{eff} \rightarrow + \infty $ happens at the area of $\kappa / M^2$ where there is no existing red line.

\subsection{Massless Particle Motion}
To discuss the motion of massless particles (light), we should modify \eqref{eq:potprob} first. Instead of expressing it in $E$ and $l$ we can use the impact parameter $b$ which,
\begin{equation}
b=\frac{l}{E} .  \label{eq:b}
\end{equation}

Modifying \eqref{eq:potprob} by \eqref{eq:b} and rescaling the affine parameters, $\tau \rightarrow \tau / E$, we get
\begin{equation} \label{eq:potprobl}
C \dot{r}^2 + V_{eff}(r) = 1
\end{equation}
where the effective potential, $V_{eff}$ is given by
\begin{equation} \label{eq:Veffl}
V_{eff}(r) = \left(\frac{b^2}{r^2}\right)\left(1 - \Delta - \frac{2M}{\sqrt{v}}\right).
\end{equation}


From the shape of the potential we know that there is no SCO or ISCO orbits. The $R_{UCO}$ for light (known as the {\it photosphere}) can be obtained easily, though not illuminating to show here. From Eq.~\eqref{eq:potprobl} we can see that when $V_{eff}(R_{UCO}) = 1$, the massless particle moves with circular orbit at $R_{UCO}$ indefinitely, while for $V_{eff}(R_{UCO}) < 1$, the particle plunges into the black hole~\cite{Sotani:2015ewa}. When $V_{eff}(R_{UCO}) > 1$, the particle is repelled before arriving at $R_{UCO}$.

\begin{figure}[h]
	\includegraphics[scale=0.7]{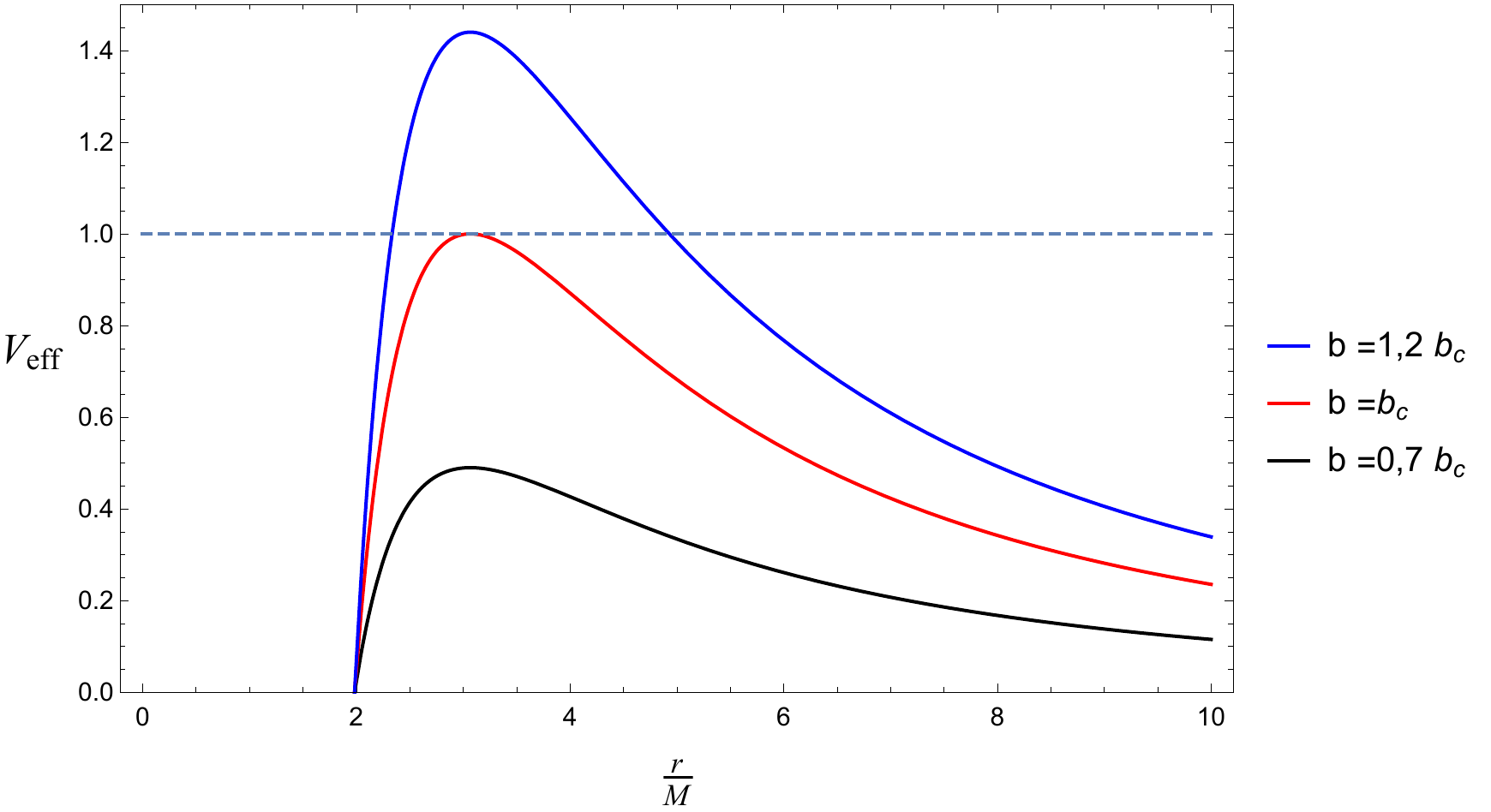}
	\caption{$V_{eff}$ for light with various value of $b$ ($M = 1$, $\Delta = 0.1$, $\kappa = 10$, and $b_c = 5.79588$).}
	\label{fig:Vlight}
\end{figure}

In Fig.~\ref{fig:Vlight} we plot $V_{eff}$ with different values of impact parameter $b$. The value of $b$ at $V_{eff}(R_{UCO}) = 1$ is defined as $b=b_c$, the critical impact parameter. In GR case (with $\Delta = 0.1$) the value is $b_c/M= 6.08581$. In EiBI, with $\Delta = 0.1$ and $\kappa = 10$ it yields $b_c/M = 5.79588$.  We can see that EiBI gives smaller value of critical impact parameter.

\subsection{Deflection of Light}

Finally, let us calculate the deflection of light that propagates near global monopole in EiBI gravity. From \eqref{eq:potprobl}, 
\begin{equation} 
\dot{r}^2  = \frac{r^2 + \Delta \kappa}{r^2}\left[ 1- \frac{b^2}{ r^2}\left(1-\Delta - \frac{2M}{\sqrt{r^2 + \Delta \kappa}}\right) \right] .
\end{equation}

When light is closest to the monopole, $\dot{r}$ becomes zero. This closest distance $r_1$ is related to the impact parameter $b$,
\begin{equation} 
b^2 = r^2_1\left(1-\Delta - \frac{2M}{\sqrt{r^2_1 + \Delta \kappa}}\right)^{-1} . \label{eq:br1relation1}
\end{equation}

Rescaling $\tau \rightarrow \tau / E$, Eq.~\eqref{eq:pl} becomes
\begin{equation}
r^2 \dot{\phi} = b.
\end{equation}
Expressing $\phi$ with respect to $r$,
\begin{eqnarray}
\left(\frac{d \phi}{dr}\right)^2 &=& \left(\frac{\dot{\phi}}{\dot{r}}\right)^2 \\
&=& \left[(r^2 + \Delta \kappa)\left( \frac{r^2}{b^2}- \left(1-\Delta - \frac{2M}{\sqrt{r^2 + \Delta \kappa}}\right) \right)\right]^{-1}.
\end{eqnarray}
We can use this relation to obtain
\begin{eqnarray}
\Delta\phi &=& 2 \int_{r_1}^{\infty} \left(\frac{\mathrm{d \phi}}{\mathrm{dr}}\right) \mathrm{dr} \\
&=& 2 \int_{r_1}^{\infty} \left[(r^2 + \Delta \kappa)\left( \frac{r^2}{b^2}- \left(1-\Delta - \frac{2M}{\sqrt{r^2 + \Delta \kappa}}\right) \right)\right]^{-1/2} dr.
\end{eqnarray}
To calculate this equation, first we use \eqref{eq:br1relation1} to set $b$
\begin{equation}
\Delta\phi = 2 \int_{r_1}^{\infty} \left[(r^2 + \Delta \kappa)\left\{ \frac{r^2}{r^2_1} \left(1-\Delta - \frac{2M}{\sqrt{r^2_1 + \Delta \kappa}}\right)- \left(1-\Delta - \frac{2M}{\sqrt{r^2 + \Delta \kappa}}\right)\right\} \right]^{-1/2}dr.
\end{equation}
We then change the variable to $u = 1/r$ and $u_1 = 1/r_1$,
\begin{equation}
\Delta\phi = 2 \int_{0}^{u_1} \left[(1 + \Delta \kappa u^2)\left\{ u^2_1 \left(1-\Delta - \frac{2M u_1}{\sqrt{1 + \Delta \kappa u_1^2}}\right)- u^2 \left(1-\Delta - \frac{2M u}{\sqrt{1 + \Delta \kappa u^2}}\right)\right\} \right]^{-1/2}du.
\end{equation}

This integral cannot be solved exactly. But we do not have to, since we are only interested in the (small) deflection angle. We instead expand the integrand to 1st order in both $\kappa$ and $M$ to get
\begin{equation}
\Delta \phi = \Delta \phi_{00} + M \Delta \phi_{01} + \kappa\Delta \phi_{10} + M \kappa\Delta \phi_{11}+{\mathcal O}(M^2\kappa^2 ),
\end{equation}
where $\Delta\phi_{01}$ is the first order expansion in  $M$ from the zeroth order expansion in $\kappa$ of $\Delta\phi$, and so on \cite{blau}. Here we look for perturbative solution in small $\kappa$ (close to GR) and $M$. Integrating the expansion yields
\begin{equation}
\Delta \phi = \frac{\pi }{\sqrt{1-\Delta }} +\frac{4 M}{b (1-\Delta )^2} -\frac{\pi  \Delta  \kappa }{4 b^2 (1-\Delta )^{3/2}} -\frac{(32-3 \pi ) \Delta  M \kappa  }{6 b^3 (1-\Delta )^3} +{\mathcal O}(M^2\kappa^2),
\end{equation}
where we use $u_1 = 1/(1-\Delta)^{1/2}b$. The deflection angle of light is then given by~
\begin{eqnarray}
\delta \phi &=& \Delta \phi - \pi \nonumber\\ 
&=& \frac{4 M}{b (1-\Delta )^2}+\pi\left(\frac{1}{\sqrt{1-\Delta }}-1\right) -{\Delta \kappa \over4b^3 \left(1-\Delta\right)^{3/2}}\left(\pi b +{2 M \left(32-3\pi\right)\over 3\left(1-\Delta\right)^{3/2}}\right)\nonumber\\
&&+{\mathcal O}(M^2\kappa^2). 
\end{eqnarray}
The first term is the deflection angle of light off a global monopole with mass $M$. The second term comes from the global monopole while excluding the effect of mass (see, for example,~\cite{Jusufi:2015laa, Jusufi:2017lsl}). It is the third term that gives contribution from the EiBI gravity in the lowest order. What is quite surprising is that this EiBI correction term is negative! This is reminiscent of the deflection angle of light off the Reissner-Nordstrom black hole, where the charge term gives negative contribution~\cite{Jusufi:2015laa, Sereno:2003nd}. The effect of (weak) EiBI gravity to the global monopole is thus to decrease its corresponding deflection angle. 

As in the case of Barriola-Vilenkin solution, ours perceived as a massive gravitating global monopole ($M\approx0$) deflects light even though exerts no gravity on its surrounding,
\begin{eqnarray}
\delta \phi&\approx&\pi\left(\frac{1}{\sqrt{1-\Delta }}-1\right) -{\Delta\kappa\pi\over4b^2\left(1-\Delta\right)^{3/2}}\nonumber\\
&\approx&4\pi^2\eta^2-{2\pi^2\eta^2\kappa\over b^2},
\end{eqnarray}
where $\Delta$ is small. If the monopole, the source of light ($S$), and the observer ($O$) are perfectly aligned then it forms an image of ring with angular diameter given by
\begin{equation}
\delta\phi\approx{4\pi^2\eta^2\ell\over\left(\ell+d\right)}-{2\pi^2\eta^2\kappa\over b^2},
\end{equation}  
with $\ell$ and $d$ are the distances of monopole to $S$ and $O$, respectively.

\section{Conclusions} \label{sec:conclusions}

To the best of our knowledge, this paper is the first one that preliminarily studies the gravitational fields and black hole solutions of global monopole in EiBI gravity, along with some of its phenomenological consequences. This nonlinear theory of gravity is parametrized by the coupling constant $\kappa$, and reduces to ordinary GR in the limit of $\kappa\rightarrow0$. We obtain the exact solutions of metric outside a global monopole in the EiBI gravity. Our solution can not be transformed into the ordinary Barriola-Vilenkin's unless in the limit of $\kappa\rightarrow0$. In that sense, our solution is distinct from them. The solution can also be perceived as an EiBI black hole eating up a global monopole. As in the case with other EiBI black hole solutions~\cite{Sotani:2014lua, Wei:2014dka, Sotani:2015ewa}, the appearance of $\kappa$ shrinks the corresponding horizon. Since there is still one horizon, we may conclude that {\it the no-hair theorem} still works in this case. It may be of some interest to investigate the EiBI black hole with noncanonical global monopole (as were studied, for example, in here~\cite{Prasetyo:2017rij}) to see whether the no-hair theorem breaks down or not. 

Studying the geodesic of massive test particles around this metric, we reveal that the ISCO is sensitively dependent on the deficit solid angle $\Delta$. There exists a maximum value of $\Delta<1$ that gives $R_{ISCO}$. Adding more monopoles renders the $R_{ISCO}$ to disappear. A peculiar behavior is given by the nonlinearity of EiBI theory. There exists a finite $\kappa$ that gives $V_{eff}$ finite at the origin. This means that at this value of $\kappa$, any particle with energy higher than the finite $V_{eff}(0)$ shall move undisturbed by the presence of monopole. For photon, on the other hand, there is no SCO or ISCO. The shape of the potential gives only one unstable extremum ($R_{UCO}$), which in the literature is known as the photosphere. When this extremum is equivalent to the (classical) photon energy, then the photon circles the monopole indefinitely. We found that the impact parameter $b$ for this to happen in EiBI is generically smaller than its GR counterpart. 

The last phenomenological issue we discuss in this paper is the (weak) deflection angle. The EiBi effect gives correction term to the angle that is actually negative. This is reminiscent to the case of light deflection angle off a Reissner-Nodrstrom black hole. Thus, the nonlinearity of this modified gravity acts like an ``effective charge" in the light deflection seen from the GR's perspective. However, we should remind the readers that our calculation was done in the weak-field limit. This result might be altered should we consider strong gravitational lensing from this monopole, as done for example in~\cite{Wei:2014dka, Sotani:2015ewa}. Another interesting possible research roadmap along this theoretical topic is by considering noncanonical global monopole in EiBI. We shall return to this issue in the forthcoming publication.

\acknowledgments

We thank Ilham Prasetyo, Jason Kristiano, Aulia Kusuma, Haryanto Siahaan and Alexander Vilenkin for useful comments. After the submission of the earlier manuscript, we were informed that Rajibul Shaikh also works on similar topic, where he previously has developed Lorentzian wormhole solutions in EiBI~\cite{Shaikh:2015oha}. We are benefitted by fruitful discussions with him. This work is partially funded by PITTA grants from Universitas Indonesia under contract no.~656/UN2.R3.1/HKP.05.00/2017 and 2267/UN2.R3.1/HKP.05.00/2018.



\begin{thebibliography}{999}

\bibitem{vilenkinshellard}
 A.~Vilenkin and E.~P.~S.~Shellard,
 {\it Cosmic Strings and other Topological Defects.} (Cambridge University Press, Cambridge, 1994).

\bibitem{Barriola:1989hx}
  M.~Barriola and A.~Vilenkin,
  ``Gravitational Field of a Global Monopole,''
  Phys.\ Rev.\ Lett.\  {\bf 63} (1989) 341.

\bibitem{Harari:1990cz}
  D.~Harari and C.~Lousto,
  ``Repulsive gravitational effects of global monopoles,''
  Phys.\ Rev.\ D {\bf 42} (1990) 2626.

\bibitem{Carames:2017ngt}
  T.~R.~P.~Carams, J.~C.~Fabris, E.~R.~Bezerra de Mello and H.~Belich,
  ``$f(R)$ global monopole revisited,''
  Eur.\ Phys.\ J.\ C {\bf 77} (2017) no.7,  496
  [arXiv:1706.02782 [gr-qc]].

\bibitem{Dadhich:1997mh}
  N.~Dadhich, K.~Narayan and U.~A.~Yajnik,
  ``Schwarzschild black hole with global monopole charge,''
  Pramana {\bf 50} (1998) 307
  [gr-qc/9703034].

\bibitem{Jin:2007fz}
  X.~h.~Jin, X.~z.~Li and D.~j.~Liu,
  ``Gravitating global k-monopole,''
  Class.\ Quant.\ Grav.\  {\bf 24} (2007) 2773
  [arXiv:0704.1685 [gr-qc]].

\bibitem{Liu:2009eh}
  D.~J.~Liu, Y.~L.~Zhang and X.~Z.~Li,
  ``A Self-gravitating Dirac-Born-Infeld Global Monopole,''
  Eur.\ Phys.\ J.\ C {\bf 60} (2009) 495
  [arXiv:0902.1051 [hep-th]].

\bibitem{Prasetyo:2017rij}
  I.~Prasetyo and H.~S.~Ramadhan,
  ``Classical defects in higher-dimensional Einstein gravity coupled to nonlinear $\sigma$-models,''
  Gen.\ Rel.\ Grav.\  {\bf 49} (2017) no.9,  115
  [arXiv:1707.06415 [gr-qc]].

\bibitem{Carames:2011xi}
  T.~R.~P.~Carames, E.~R.~Bezerra de Mello and M.~E.~X.~Guimaraes,
  ``On the motion of a test particle around a global monopole in a modified gravity,''
  Mod.\ Phys.\ Lett.\ A {\bf 27} (2012) 1250177
  [arXiv:1111.1856 [gr-qc]].

\bibitem{Carames:2011uu}
  T.~R.~P.~Carames, E.~R.~Bezerra de Mello and M.~E.~X.~Guimaraes,
  ``Gravitational Field of a Global Monopole in a Modified Gravity,''
  Int.\ J.\ Mod.\ Phys.\ Conf.\ Ser.\  {\bf 3} (2011) 446
  [arXiv:1106.4033 [gr-qc]].

\bibitem{Man:2013sf}
  J.~Man and H.~Cheng,
  ``Thermodynamic quantities of a black hole with an f(R) global monopole,''
  Phys.\ Rev.\ D {\bf 87} (2013) no.4,  044002
  [arXiv:1301.2739 [hep-th]].

\bibitem{Lustosa:2015hwa}
  F.~B.~Lustosa, M.~E.~X.~Guimar‹es, C.~N.~Ferreira and J.~L.~Neto,
  ``Thermodynamical Analysis of a Black Hole with a Global Monopole Within a Class of a f(R) Gravity,''
  arXiv:1510.08176 [hep-th].

\bibitem{MoraisGraca:2012sq}
  J.~P.~Morais Graca and V.~B.~Bezerra,
  ``Gravitational field of a rotating global monopole in f(R) theory,''
  Mod.\ Phys.\ Lett.\ A {\bf 27} (2012) 1250178.

\bibitem{Man:2012fa}
  J.~Man and H.~Cheng,
  ``Analytical discussion on strong gravitational lensing for a massive source with a f(R) global monopole,''
  Phys.\ Rev.\ D {\bf 92} (2015) no.2,  024004
  [arXiv:1205.4857 [gr-qc]].

\bibitem{Banados:2010ix}
  M.~Banados and P.~G.~Ferreira,
  ``Eddington's theory of gravity and its progeny,''
  Phys.\ Rev.\ Lett.\  {\bf 105} (2010) 011101
   Erratum: [Phys.\ Rev.\ Lett.\  {\bf 113} (2014) no.11,  119901]
  [arXiv:1006.1769 [astro-ph.CO]].

\bibitem{Delsate:2012ky}
  T.~Delsate and J.~Steinhoff,
  ``New insights on the matter-gravity coupling paradigm,''
  Phys.\ Rev.\ Lett.\  {\bf 109} (2012) 021101
  [arXiv:1201.4989 [gr-qc]].

\bibitem{Pani:2012qd}
  P.~Pani and T.~P.~Sotiriou,
  ``Surface singularities in Eddington-inspired Born-Infeld gravity,''
  Phys.\ Rev.\ Lett.\  {\bf 109} (2012) 251102
  [arXiv:1209.2972 [gr-qc]].

\bibitem{Sotani:2014goa}
  H.~Sotani,
  ``Observational discrimination of Eddington-inspired Born-Infeld gravity from general relativity,''
  Phys.\ Rev.\ D {\bf 89} (2014) no.10,  104005
  [arXiv:1404.5369 [astro-ph.HE]].

\bibitem{Avelino:2012ge}
  P.~P.~Avelino,
  ``Eddington-inspired Born-Infeld gravity: astrophysical and cosmological constraints,''
  Phys.\ Rev.\ D {\bf 85} (2012) 104053
  [arXiv:1201.2544 [astro-ph.CO]].

\bibitem{Prasetyo:2017hrb}
  I.~Prasetyo, I.~Husin, A.~I.~Qauli, H.~S.~Ramadhan and A.~Sulaksono,
  ``Neutron stars in the braneworld within the Eddington-inspired Born-Infeld gravity,''
  arXiv:1708.04837 [astro-ph.CO].

\bibitem{Qauli:2016vza}
  A.~I.~Qauli, M.~Iqbal, A.~Sulaksono and H.~S.~Ramadhan,
  ``Hyperons in neutron stars within an Eddington-inspired Born-Infeld theory of gravity,''
  Phys.\ Rev.\ D {\bf 93} (2016) no.10,  104056
  [arXiv:1605.01152 [astro-ph.SR]].

\bibitem{Qauli:2017ntr}
  A.~I.~Qauli, A.~Sulaksono, H.~S.~Ramadhan and I.~Husin,
  ``Apparent equation of state of compact stars within the Eddington-inspired Born-Infeld theory,''
  arXiv:1710.03988 [gr-qc].

\bibitem{BeltranJimenez:2017doy}
  J.~Beltran Jimenez, L.~Heisenberg, G.~J.~Olmo and D.~Rubiera-Garcia,
  ``Born-Infeld inspired modifications of gravity,''
  arXiv:1704.03351 [gr-qc].

\bibitem{Sotani:2014lua}
  H.~Sotani and U.~Miyamoto,
  ``Properties of an electrically charged black hole in Eddington-inspired Born-Infeld gravity,''
  Phys.\ Rev.\ D {\bf 90} (2014) 124087
  [arXiv:1412.4173 [gr-qc]].

\bibitem{Wei:2014dka}
  S.~W.~Wei, K.~Yang and Y.~X.~Liu,
  ``Black hole solution and strong gravitational lensing in Eddington-inspired Born-Infeld gravity,''
  Eur.\ Phys.\ J.\ C {\bf 75} (2015) 253
   Erratum: [Eur.\ Phys.\ J.\ C {\bf 75} (2015) 331]
  [arXiv:1405.2178 [gr-qc]].

\bibitem{Deser:1998rj}
  S.~Deser and G.~W.~Gibbons,
  ``Born-Infeld-Einstein actions?,''
  Class.\ Quant.\ Grav.\  {\bf 15} (1998) L35
  [hep-th/9803049].

\bibitem{Vollick:2003qp}
  D.~N.~Vollick,
  ``Palatini approach to Born-Infeld-Einstein theory and a geometric description of electrodynamics,''
  Phys.\ Rev.\ D {\bf 69} (2004) 064030
  [gr-qc/0309101].

\bibitem{Tan:2017egu}
  H.~Tan, J.~Yang, J.~Zhang and T.~He,
  ``The global monopole spacetime and its topological charge,''
  arXiv:1705.00817 [gr-qc].

\bibitem{Garfinkle:1990qj}
  D.~Garfinkle, G.~T.~Horowitz and A.~Strominger,
  ``Charged black holes in string theory,''
  Phys.\ Rev.\ D {\bf 43} (1991) 3140
   Erratum: [Phys.\ Rev.\ D {\bf 45} (1992) 3888].

\bibitem{proceding}
R.~D.~Lambaga and H.~S.~Ramadhan,
``Massive Particle around Black Hole with Global Monopole
in Eddington-inspired Born-Infeld Gravity,''
presented at the International Symposium on Current Progress in Mathematics and Sciences (ISCPMS) 2017, Bali, Indonesia.

\bibitem{Sotani:2015ewa}
  H.~Sotani and U.~Miyamoto,
  ``Strong gravitational lensing by an electrically charged black hole in Eddington-inspired Born-Infeld gravity,''
  Phys.\ Rev.\ D {\bf 92} (2015) no.4,  044052
  [arXiv:1508.03119 [gr-qc]].

\bibitem{blau}
M.~Blau,
``Lecture Notes on General Relativity,"
http://www.blau.itp.unibe.ch/Lecturenotes.html

\bibitem{Jusufi:2015laa}
  K.~Jusufi,
  ``Gravitational lensing by Reissner-Nordstršm black holes with topological defects,''
  Astrophys.\ Space Sci.\  {\bf 361} (2016) no.1,  24
  [arXiv:1510.08526 [gr-qc]].

\bibitem{Jusufi:2017lsl}
  K.~Jusufi, M.~C.~Werner, A.~Banerjee and A.~Ovgun,
  ``Light Deflection by a Rotating Global Monopole Spacetime,''
  Phys.\ Rev.\ D {\bf 95} (2017) no.10,  104012
  [arXiv:1702.05600 [gr-qc]].

\bibitem{Sereno:2003nd}
  M.~Sereno,
  ``Weak field limit of Reissner-Nordstrom black hole lensing,''
  Phys.\ Rev.\ D {\bf 69} (2004) 023002
  doi:10.1103/PhysRevD.69.023002
  [gr-qc/0310063].

\bibitem{Shaikh:2015oha}
  R.~Shaikh,
  ``Lorentzian wormholes in Eddington-inspired Born-Infeld gravity,''
  Phys.\ Rev.\ D {\bf 92} (2015) 024015
  [arXiv:1505.01314 [gr-qc]].

  

 
  
\end{thebibliography}
\end{document}